\documentclass{aa}

\usepackage{graphicx}

\usepackage{txfonts}

\usepackage{xcolor}
\usepackage[utf8]{inputenc}

\definecolor{xlinkcolor}{cmyk}{1,0.6,0,0}

\usepackage[breaklinks=true, 
    bookmarks=false,         
     pdfnewwindow=true,      
     colorlinks=true,    
     linkcolor=xlinkcolor,     
     citecolor=xlinkcolor,     
     filecolor=xlinkcolor,  
     urlcolor=xlinkcolor,      
final=true
]{hyperref}

\begin{document}

   \title{The habitability trade-off: Chemical decoupling and quenching in massive galaxies}

   \subtitle{}

   \author{Ana Mitra{\v s}inovi{\' c}\inst{1}
            \and
           Nata{\v s}a Pavlov\inst{2}
           \and
           Branislav Vukoti{\' c}\inst{1}
            \and
           Stanislav Milo{\v s}evi{\' c}\inst{2}
         }

   \institute{$^1$ Astronomical Observatory, Volgina 7, 11060 Belgrade, Serbia\\
              \email{amitrasinovic@aob.rs}\\
              $^2$ Faculty of Mathematics, University of Belgrade, Studentski trg 16, 11158 Belgrade, Serbia\\
              }

   \date{Received ; accepted }

  \abstract
  {Massive galaxies experience complex evolutionary processes, including mergers and gas accretion, which can disrupt the chemical equilibrium between their stellar and gaseous components. Using the IllustrisTNG (TNG100) simulation at $z=0$, we investigated the prevalence and physical properties of such chemically decoupled systems within the massive star-forming galaxy population. We identify a substantial subpopulation ($\sim 31.5\%$ of the sample) that exhibits systematic stellar-gas decoupling, characterised by a metal-rich stellar component coexisting with a diluted gas reservoir. These non-equilibrium galaxies are closely linked to recent merger activity and partial quenching, and display systematically suppressed star-formation rates and reduced gas fractions, consistent with a transitional evolutionary phase. We then examined the implications of this phase for galaxy-scale habitability prescriptions by applying a terrestrial planet abundance proxy that combines stellar mass, gas-phase metallicity, and the rate of sterilising events. Despite their diluted gas reservoirs, non-equilibrium galaxies dominate the high end of the inferred present-day habitability proxy distribution, exceeding equilibrium systems by more than an order of magnitude. We interpret this as a habitability trade-off: the same gas dilution and quenching processes that reduce the efficiency of future terrestrial planet formation simultaneously create a transient phase of suppressed radiation hazards for existing planets. The Andromeda galaxy (M31) shows qualitative similarities to this chemically decoupled population, suggesting that galaxies exiting their peak star-forming phase represent a distinct and highly relevant demographic for galaxy-scale habitability. Galactic habitability is therefore intrinsically time-dependent.
  }

   \maketitle

\section{Introduction}

The evolution of massive galaxies is governed by a complex interplay of gas accretion, star formation, and feedback, all of which leave distinct imprints on their chemical abundances. In the local Universe, tight scaling relations between the stellar mass, star-formation rate (SFR), and gas-phase metallicity, collectively referred to as the fundamental plane of galaxy formation \citep{Mannucci+2010}, indicate that galaxies typically evolve in a state of approximate chemical equilibrium. Under these conditions, the metallicity of the established stellar population broadly traces the chemical environment of the star-forming interstellar medium. This equilibrium picture is supported by the tight mass–metallicity relations (MZRs) observed for both stars and gas \citep[e.g.][]{Tremonti+2004, Gallazzi+2005, Maiolino+Mannucci2019}.

Massive galaxies, however, frequently experience mergers, gas inflows, and quenching events that can disrupt this balance. Inflows of metal-poor gas can dilute the interstellar medium while leaving the pre-existing stellar population largely unchanged, resulting in a measurable offset between stellar and gas-phase metallicities \citep[e.g.][]{Yates+2012MNRAS.422..215Y, Yates+2014MNRAS.439.3817Y, Bustamante+2018MNRAS.479.3381B, Sparre+2022MNRAS.509.2720S, perez-diaz_departure_2024, Pan+2025ApJ...982..130P}. In addition, terminal explosions in multiple stellar systems can alter stellar orbits and contribute to the redistribution of chemically enriched stars to the galactic outskirts \citep{article_tutukov2025}. In such situations, stellar mass and stellar metallicity may no longer provide a reliable description of the chemical conditions that govern ongoing or future star formation.

Departures from chemical equilibrium are relevant not only for galaxy evolution, but also for statistical frameworks that connect global galaxy properties to the likelihood of them forming and sustaining habitable planets \citep[e.g.][]{Lineweaver2001, Fischer+Valenti2005ApJ...622.1102F, Dayal+2015ApJ...810L...2D}. Habitability has traditionally been studied at the scale of individual planetary systems, with a focus on identifying Earth-like planets and the conditions required for life \citep[e.g.][]{Hart1979Icar...37..351H, Kasting+1993Icar..101..108K, Seager2013Sci...340..577S}. As observational capabilities and theoretical models have matured, this perspective has expanded to include the galactic environments in which planetary systems reside \citep{Gonzalez+2001, Lineweaver+2004, Balbi+Tombesi2017NatSR...716626B, Kaib2018haex.bookE..59K}. Because exoplanet detections remain strongly biased towards the solar neighbourhood \citep[see e.g.][]{Maliuk+Budaj2020A&A...635A.191M}, assessments of habitability beyond the Milky Way (MW) necessarily rely on indirect approaches that link large-scale galaxy properties to planet formation and survival probabilities.

The notion that habitability can vary across different galactic environments has been explored extensively through the concept of the galactic habitable zone (GHZ). First formalised for the MW by \citet{Gonzalez+2001} and further developed by \citet{Lineweaver+2004}, the GHZ framework relates metallicity, star formation, and energetic feedback to the spatial and temporal distribution of potentially habitable environments. Subsequent studies have refined this picture using increasingly detailed observational constraints, largely focused on the MW \citep[e.g.][]{Prantzos2008, Gowanlock+2011, 2016A&A...592A..96G, Spitoni+2017, Gowanlock+gordon_habitability_2018, Kokaia+Davies2019, Spinelli+2021, 2023PASA...40...54M}. Similar analyses have been extended to nearby massive galaxies, most notably the Andromeda galaxy \citep[M31; e.g.][]{Carigi+2013, Spitoni+2014}.

Metallicity plays a central role in these frameworks because of its observed correlation with planet formation \citep{Lineweaver2001}. Giant planets are more common around metal-rich stars \citep[e.g.][]{10.1093/mnras/285.2.403,2001A&A...373.1019S, Fischer+Valenti2005ApJ...622.1102F, 2005A&A...437.1127S}, while smaller planets appear preferentially around higher-metallicity stars, although with a weaker dependence \citep[e.g.][]{2015ApJ...808..187B, Wang+Fischer2015AJ....149...14W, 2019ApJ...873....8Z, 2020AJ....160..253L}. The SFR is also commonly employed in galaxy-scale habitability models, as it governs both the emergence of new planetary systems and the frequency of potentially sterilising events \citep[e.g.][]{Lineweaver+2004, Gowanlock+2011, Dayal+2015ApJ...810L...2D}. Elevated star formation increases supernova and gamma-ray burst rates, whereas suppressed star formation reduces these hazards but limits future planet formation.

Building on these ideas, several studies have evaluated habitability on galactic scales. \citet{Suthar+McKay2012} applied a metallicity-based methodology to elliptical galaxies, arguing that diverse morphological types may host extended habitable regions. Later work suggested that massive, spheroid-dominated systems provide particularly favourable environments for terrestrial planets \citep{Dayal+2015ApJ...810L...2D, Zackrisson+2016}, although alternative interpretations have been proposed \citep[e.g.][]{Whitmire2020}. In parallel, cosmological simulations have become an indispensable tool for tracking galaxy evolution (for reviews of the ecosystem of cosmological simulations, see \citealt{Vogelsberger+2020NatRP...2...42V} and \citealt{Crain+vaddeVoort2023ARA&A..61..473C}) and are increasingly adopted in galactic habitability studies \citep[e.g.][]{Vukotic+2016, Forgan+2017, Stanway+2018, Stojkovic+2019MNRAS, Stojkovic+2019SerAJ}. These numerical laboratories enable the coupled evolution of baryons and dark matter to be followed self-consistently, providing a framework in which chemical enrichment, star formation, and hierarchical assembly can be examined together.

In our recent work \citep{2025PASA...42...66M}, we revisited claims of bimodality in galactic habitability \citep{Stojkovic+2019MNRAS}, focusing on the low-mass regime and a population of dwarf galaxies with enhanced metallicities relative to the MZR. During that analysis, we identified a substantial population of massive galaxies exhibiting a marked lack of correlation between their gas-phase and stellar metallicities. Although this chemically decoupled population was beyond the scope of the previous study, its presence raises important questions regarding the validity of equilibrium-based assumptions in galaxy-scale habitability prescriptions.

In this study we focused explicitly on these massive, chemically decoupled galaxies. Using the IllustrisTNG simulation at $z = 0$, we characterised their chemical properties, star-formation activity, and gas content, and assessed whether stellar–gas metallicity decoupling represents a distinct evolutionary phase among massive systems. We then examined the implications of this decoupling for commonly used galaxy-scale habitability metrics, emphasising that such interpretations must be considered within the broader context of galaxy evolution.

This paper is organised as follows. Section~\ref{sec:methods} describes the simulation data, methodology, and sample selection. Section~\ref{sec:phys-results} presents the chemical and star-formation properties of non-equilibrium and equilibrium galaxies. Section~\ref{sec:astrobio-results} explores the implications of chemical decoupling for galaxy-scale habitability. Section~\ref{sec:discussion} discusses the physical interpretation of our results in an observational context, and Sect.~\ref{sec:conslusions} summarises our conclusions.

\section{Methods and data}\label{sec:methods}

In this work we utilised the IllustrisTNG project\footnote{Publicly available at \url{https://www.tng-project.org/data/}}, a suite of state-of-the-art cosmological magnetohydrodynamic simulations \citep{TNGmethods2017, TNGmethods2018, Nelson+2019ComAC}. The simulations were evolved using the \texttt{Arepo} moving-mesh code \citep{Springel2010AREPO}, which couples a robust galaxy formation model with accurate hydrodynamics to reproduce key observational scaling relations across a broad range of redshifts. The simulation suite consists of three main cubic volumes\footnote{Excluding the recently released TNG-Cluster \citep{Nelson+2024-tngcluster}.}, which differ in particle and spatial resolution (driven by the adopted softening length parameter). Since they also differ in total volume, the number of galaxies also differs. These differences drive the choice of a simulation box for research purposes. We used the TNG100 simulation \citep{Marinacci+2018, Naiman+2018, Nelson+2018, Pillepich+2018, Springel+2018}, which has a volume of $(110.7\;\mathrm{Mpc})^3$ and a baryonic mass resolution of approximately $1.4\times10^6\;\mathrm{M_\odot}$. This volume offers an optimal balance between statistical sampling of massive galaxies and the mass and spatial resolution necessary to resolve internal galactic properties.

We constructed a parent sample of galaxies from the redshift $z=0$ snapshot. To ensure that our analysis is restricted to well-resolved, physical systems of cosmological origin, we applied a series of standard filtering criteria. We selected only those structures flagged as genuine galaxies\footnote{See the detailed discussion in \citet{2025PASA...42...66M}.} that contain a non-zero dark matter component and both stellar and gaseous components, sufficiently and robustly resolved. In practice, this means that we required galaxies to be resolved with a sufficient number of stellar particles and gas cells for integrated metallicities and SFRs to be meaningful \citep[see][]{Onions+2012MNRAS.423.1200O}. We also restricted the sample to systems with non-zero gas-phase metallicity, ensuring that we only analysed systems that have participated in at least some degree of chemical enrichment. For all gas-related quantities, including gas-phase metallicity, we considered all gravitationally bound gas associated with each subhalo, irrespective of temperature or star-formation state, rather than restricting the analysis to star-forming gas only.

To quantify the chemical evolution of our sample, we analysed the MZR for both the stellar and gas-phase components. We adopted the empirical asymptotic formalism proposed by \citet{Zahid2014ApJ...791..130Z}, which characterises the MZR as a power-law rise at low masses that saturates to a constant metallicity at high masses, as a function of stellar mass. The relation is given by

\begin{equation}
    Z(M_\star) = Z_0 - \log \left[ 1 + \left(\frac{M_0}{M_\star}\right)^\gamma \right]
,\end{equation}

\noindent where $M_\star$ is the stellar mass, $Z(M_\star)$ is the total metallicity, and the free parameters are the asymptotic metallicity limit (saturation level; $Z_0$), the characteristic turnover mass above which the metallicity saturates ($M_0$), and the power-law slope at the low-mass end ($\gamma$).

We performed independent fits of this function to the stellar and gas-phase components of our parent sample. It is important to note that while observational studies typically rely on specific emission lines (e.g. strong-line diagnostics from HII regions) to estimate metallicities \citep{Tremonti+2004, Kewley+Ellison2008ApJ...681.1183K, Maiolino+Mannucci2019}, our analysis utilises the total intrinsic metallicity directly available from the simulation. Furthermore, for the gas-phase component, we included all gravitationally bound gas associated with the subhalo, consistent with the definition adopted above. Consequently, while our derived structural parameters ($M_0$ and $\gamma$) can be broadly compared to observational results \citep[e.g.][]{Tremonti+2004, Zahid2014ApJ...791..130Z}, normalisation ($Z_0$) reflects the total metal budget of the simulation rather than a specific calibration scale.

\begin{figure}[!h]
\centering \includegraphics[width=\columnwidth, keepaspectratio]{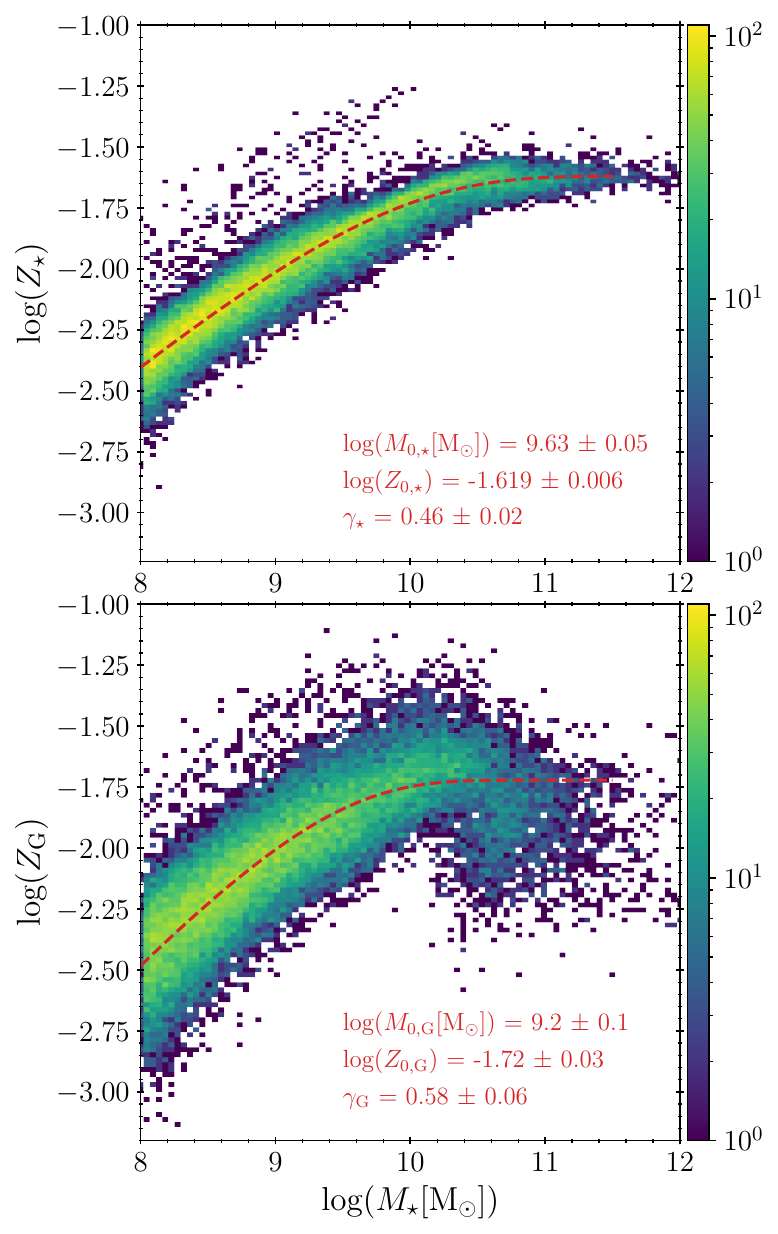}
\caption{Stellar (top) and gas-phase (bottom) MZRs for the TNG100 parent sample at $z=0$. The colour map indicates the logarithmic number density of galaxies. Dashed red curves show the best-fitting asymptotic relations obtained by fitting the \cite{Zahid2014ApJ...791..130Z} parameterisation to the TNG100 data, with the resulting fit parameters ($M_0$,$Z_0$,$\gamma$) displayed in the corresponding panels.}
\label{fig:MZRfits}
\end{figure}

Figure~\ref{fig:MZRfits} presents the resulting distributions and the best-fitting relations, with fitting parameters included in the panels. The derived parameters confirm that the gas-phase and stellar components follow distinct evolutionary tracks, necessitating the independent fitting approach employed here. In other words, we find a clear structural difference between the two components. The gas-phase relation saturates at a lower mass, the gas-phase turnover mass $\log(M_{0,G}/\mathrm{M}_\odot) = 9.2 \pm 0.1$, compared to the stellar component, with the stellar turnover mass $\log(M_{0,\star}/\mathrm{M}_\odot) = 9.63 \pm 0.05$. The gas-phase best-fit parameters are broadly consistent with the seminal Sloan Digital Sky Survey (SDSS) results of \citet{Tremonti+2004} and \citet{Zahid2014ApJ...791..130Z}. The turnover mass is at the lower end of the observationally inferred range, while the low-mass slope, $\gamma_G = 0.59 \pm 0.06$, is fully consistent with the values reported in observational studies. The modest offset in turnover mass is expected, given the differences in metallicity definition and gas selection, since our analysis includes all bound gas rather than only star-forming regions, which are typically probed by nebular emission lines. Furthermore, we recover the observational trend, where the low-mass power-law slope is shallower for stars ($\gamma_\star = 0.46 \pm 0.02$) than for gas ($\gamma_G = 0.59 \pm 0.06$), reflecting the different timescales probed by these components. The gas phase is sensitive to instantaneous baryon cycle processes, while the stars represent an integrated record of enrichment \citep[e.g.][]{Gallazzi+2005, Zahid2014ApJ...791..130Z}.

Crucially, the asymptotic metallicity limits differ by $\sim 0.1\;\mathrm{dex}$ (stellar $\log Z_{0,\star}$ = -1.62 versus gas-phase $\log Z_{0,G} = -1.72$). This offset suggests that, in the high-mass regime, the gas phase is systematically more metal-poor than the saturated stellar population. For the remainder of this work, we imposed a lower stellar-mass limit of $\log(M_\star/\mathrm{M}_\odot) \geq 10$, which approximately corresponds to the mass scale above which chemically decoupled galaxies begin to emerge in the simulation. That is, above that mass, the gas-phase metallicity has a significant and systematic departure from MZR (see Fig.~\ref{fig:MZRfits}). To identify galaxies that exhibit chemical decoupling between their stellar and gaseous components, we defined a metallicity-ratio diagnostic based on the asymptotic limits of the fitted MZRs. Specifically, we computed the stellar-to-gas metallicity ratio, $Z_\star/Z_\mathrm{G}$, for all galaxies in the mass-selected sample. We then adopted the best-fitting asymptotic metallicity ratio, $Z_{0,\star}/Z_\mathrm{0, G}$, as an operational reference value for the chemical equilibrium. Galaxies with stellar-to-gas metallicity ratios below this threshold are classified as (chemically) equilibrium systems, while those above it are classified as (chemically) non-equilibrium systems. The exact value of this threshold (used for classification and derived from the ratio of asymptotic metallicity limits) is $\log(Z_{0,\star}/Z_\mathrm{0, G}) = 0.1$. Individual galaxies are then classified based on their measured values of $\log(Z_\star/Z_\mathrm{G})$ relative to this reference.

For the remainder of this work, we also restricted our analysis to galaxies with non-zero SFRs, excluding fully quenched systems. This selection is motivated by two considerations. First, we aimed to characterise chemical decoupling in galaxies that remain at least weakly star-forming, where gas-phase enrichment and dilution processes are still ongoing. Second, the habitability framework adopted in Sect.~\ref{sec:astrobio-results} \citep[i.e.][]{Dayal+2015ApJ...810L...2D} explicitly depends on the SFR (and diverges for completely quenched galaxies) and is therefore most naturally applied to galaxies with measurable star-formation activity. The final sample thus consists of massive, star-forming galaxies that are subdivided into equilibrium and non-equilibrium populations according to their stellar-to-gas metallicity ratios.

\section{Chemical decoupling and star formation in massive galaxies}\label{sec:phys-results}

Following the methodology and sample definition described in Sect.~\ref{sec:methods}, our final mass-selected star-forming galaxy sample at $z=0$ consists of 4718 systems. Of these, 3231 galaxies are classified as (chemically) equilibrium and 1487 as (chemically) non-equilibrium according to the stellar-to-gas metallicity ratio criterion introduced above. Non-equilibrium galaxies therefore constitute approximately 31.5 per cent of the massive, star-forming population considered in this work. In the remainder of this section, we examine the chemical, star-formation, and gas-content properties of these two populations.

\subsection{Equilibrium and non-equilibrium galaxies}\label{sec:equil_and_non_equil}

\begin{figure}[!h]
\centering \includegraphics[width=\columnwidth, keepaspectratio]{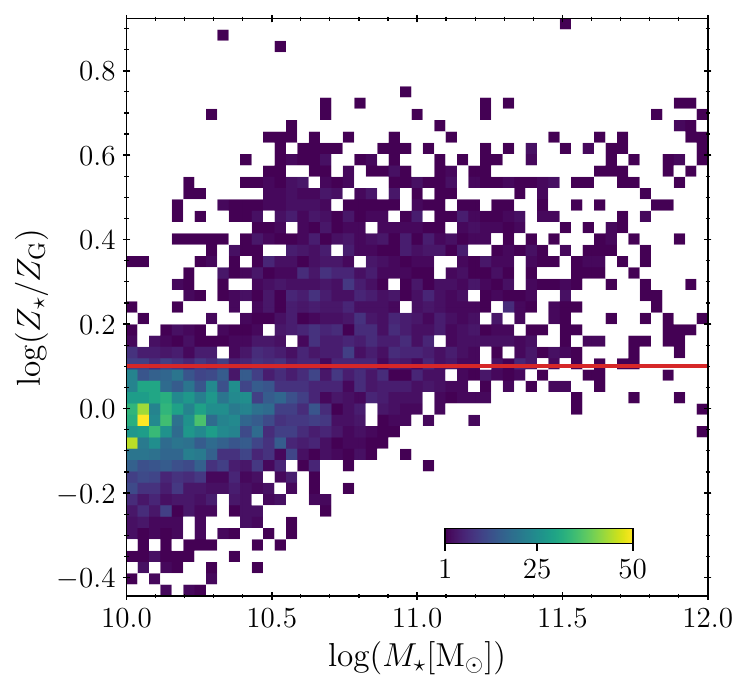}
\caption{Stellar-to-gas metallicity ratio as a function of stellar mass for the mass-selected galaxy sample. The horizontal solid red line marks the ratio of asymptotic metallicities inferred from the stellar and gas-phase MZR fits. This boundary is used to separate (chemically) equilibrium and (chemically) non-equilibrium galaxies. The colour map indicates the number of galaxies per bin.}
\label{fig:MainLogRatio}
\end{figure}

Figure~\ref{fig:MainLogRatio} illustrates the distribution of the stellar-to-gas metallicity ratio as a function of stellar mass for the massive galaxy sample. Although the majority of galaxies exhibit comparable enrichment levels in their stellar and gaseous components, a substantial population occupies a regime of elevated metallicity ratio $Z_\star/Z_\mathrm{G}$, corresponding to galaxies in which the stellar metallicity significantly exceeds the present-day gas-phase metallicity.

The emergence of this chemically decoupled population (i.e. non-equilibrium galaxies) is most prominent at high stellar masses, consistent with the mass scale at which the gas-phase MZR begins to significantly deviate from the stellar relation. This behaviour suggests that non-equilibrium galaxies are not random outliers, but rather constitute a systematic subpopulation of massive systems in which the enrichment histories of stars and gas have become decoupled.

A natural interpretation is that the stellar metallicity in these galaxies reflects an integrated record of past chemical enrichment, whereas the present-day gas reservoir has experienced dilution or incomplete re-enrichment. As mentioned previously, such a configuration can arise through external accretion of relatively metal-poor gas, merger-driven inflows, or other late evolutionary processes that affect the interstellar medium more strongly than the established stellar component. Similar mechanisms have been proposed in both semi-analytic and hydrodynamical studies of metallicity dilution following gas accretion and mergers \citep[e.g.][]{Yates+2012MNRAS.422..215Y, Yates+2014MNRAS.439.3817Y, Bustamante+2018MNRAS.479.3381B}. 

\begin{figure}[!h]
\centering \includegraphics[width=\columnwidth, keepaspectratio]{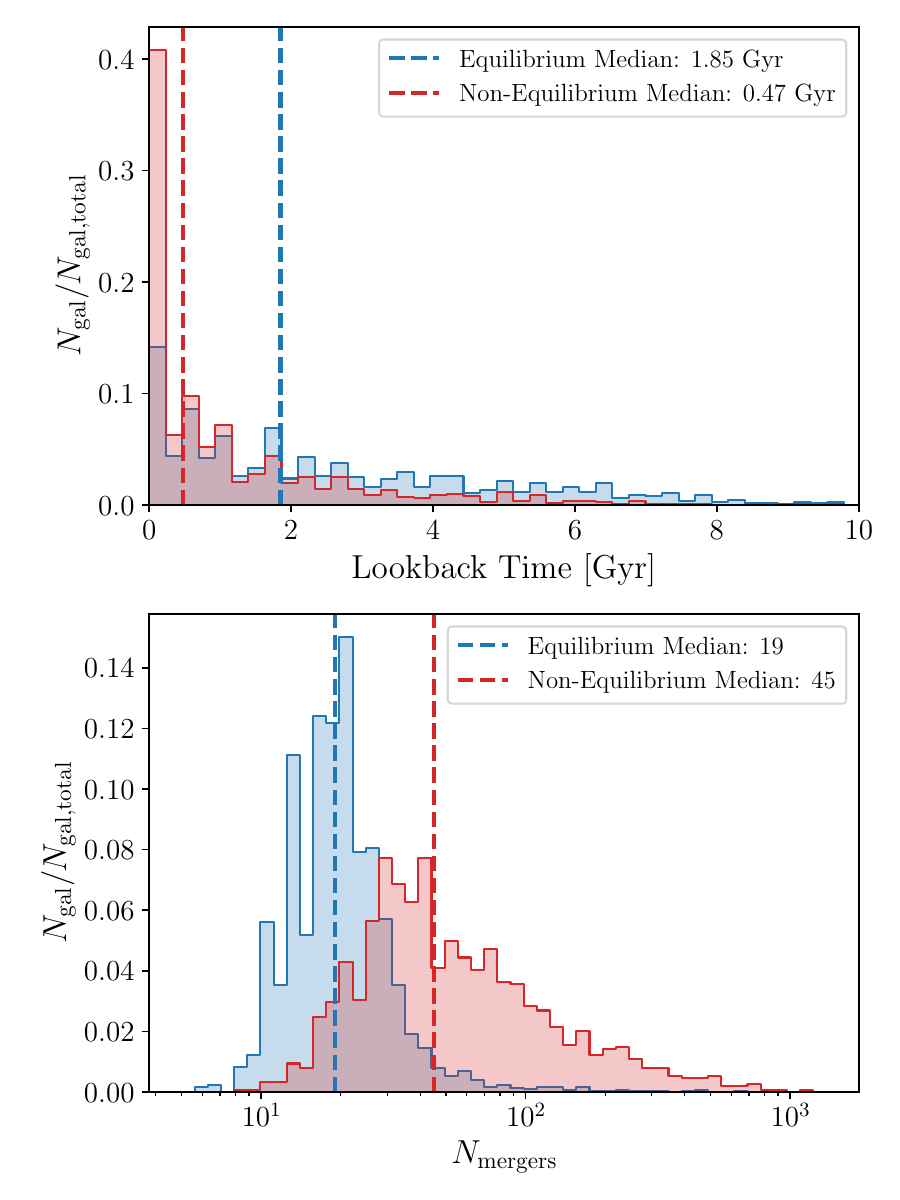}
\caption{Merger statistics for the equilibrium (blue) and non-equilibrium (red) galaxy populations, normalised by the total number of galaxies in each population. Top panel: Distribution of lookback times to the most recent merger event of any stellar mass ratio, as recorded in the supplementary merger history catalogue \citep{Rodriguez-Gomez+2017MNRAS.467.3083R, Eisert+2023MNRAS.519.2199E}. Bottom panel: Distribution of the total number of mergers experienced over the full assembly history of each galaxy. In both panels, vertical dashed lines indicate the median values for each population, with exact values given in the legends.}
\label{fig:mergers}
\end{figure}

To further assess whether chemical decoupling is linked to recent interaction activity, we examined merger histories using the official IllustrisTNG supplementary \texttt{Merger History} catalogue \citep{Rodriguez-Gomez+2017MNRAS.467.3083R, Eisert+2023MNRAS.519.2199E}, which provides lookback times to the most recent merger event and the most recent major merger for each subhalo, among other information. Chemically decoupled galaxies exhibit systematically more recent merger activity than their counterparts. For major mergers (defined as those with stellar mass ratio $>1/4$), the chemically decoupled population has a median lookback time of $8.23\;\mathrm{Gyr}$, compared to $9.90\;\mathrm{Gyr}$ for equilibrium galaxies. A two-sample Kolmogorov–Smirnov test confirms that the two distributions are statistically distinct ($D = 0.148$, $p \simeq 2 \times 10^{-19}$). The contrast is even stronger when considering mergers of any mass ratio: chemically decoupled galaxies have a median time since the last merger of only $0.47\;\mathrm{Gyr}$, compared to $1.85\; \mathrm{Gyr}$ for equilibrium systems, with the KS test again indicating a highly significant difference ($D = 0.317$, $p \simeq 3 \times 10^{-91}$). However, the similarity in both the number of major mergers and their corresponding lookback times between the two populations suggests that major mergers alone are unlikely to be the primary driver of the present-day chemical decoupling. In particular, the median total number of major mergers is $3$ for equilibrium galaxies and $4$ for non-equilibrium systems, a difference that is modest in absolute terms and consistent with the relatively long median lookback times to the last major merger reported above ($9.90\;\mathrm{Gyr}$ and $8.23\;\mathrm{Gyr}$ respectively), which place the most recent major merger events at epochs well before the present day for both populations. This indicates that additional processes must play a more important role in shaping the observed differences.

To further investigate this, we considered the full merger histories of the two populations, including mergers of all mass ratios. In Fig.~\ref{fig:mergers} we show the normalised distributions of the lookback time to the most recent merger event and of the total number of mergers throughout the entire assembly history, for both populations. In contrast to the behaviour seen for major mergers, non-equilibrium galaxies exhibit systematically more recent merger activity, with a median lookback time of $0.47\;\mathrm{Gyr}$ compared to $1.85\;\mathrm{Gyr}$ for equilibrium systems, as reported above and indicated in the legend of Fig.~\ref{fig:mergers}. The contrast in recent merger activity is also visually striking: the non-equilibrium population is strongly concentrated towards very recent events, with its distribution peaking well below $1\;\mathrm{Gyr}$, whereas the equilibrium population is spread more uniformly across the available lookback time. This reinforces the picture suggested by the median values alone and demonstrates that recent accretion or interaction activity is a systematic feature of the chemically decoupled population rather than a property of a small subset of outliers. In addition, the total merger counts across all mass ratios differ substantially, with median values of $19$ and $45$ for equilibrium and non-equilibrium galaxies, respectively. This points to a picture in which the cumulative effect of minor mergers and gas accretion events (individually less disruptive but collectively more recent and more frequent) drives and sustains the dilution of the gas-phase metallicity in chemically decoupled systems. Such processes are expected to introduce metal-poor gas into the interstellar medium, leading to the dilution of the gas-phase metallicity while leaving the pre-existing stellar component largely unaffected. This provides a natural explanation for the observed offset between stellar and gas-phase metallicities and supports a scenario in which chemical decoupling arises primarily from recent, lower-mass interactions rather than from major merger events.

\subsection{Star-formation suppression in chemically decoupled systems}\label{sec:SFRsupression}

\begin{figure*}[!h]
\centering \includegraphics[width=2\columnwidth, keepaspectratio]{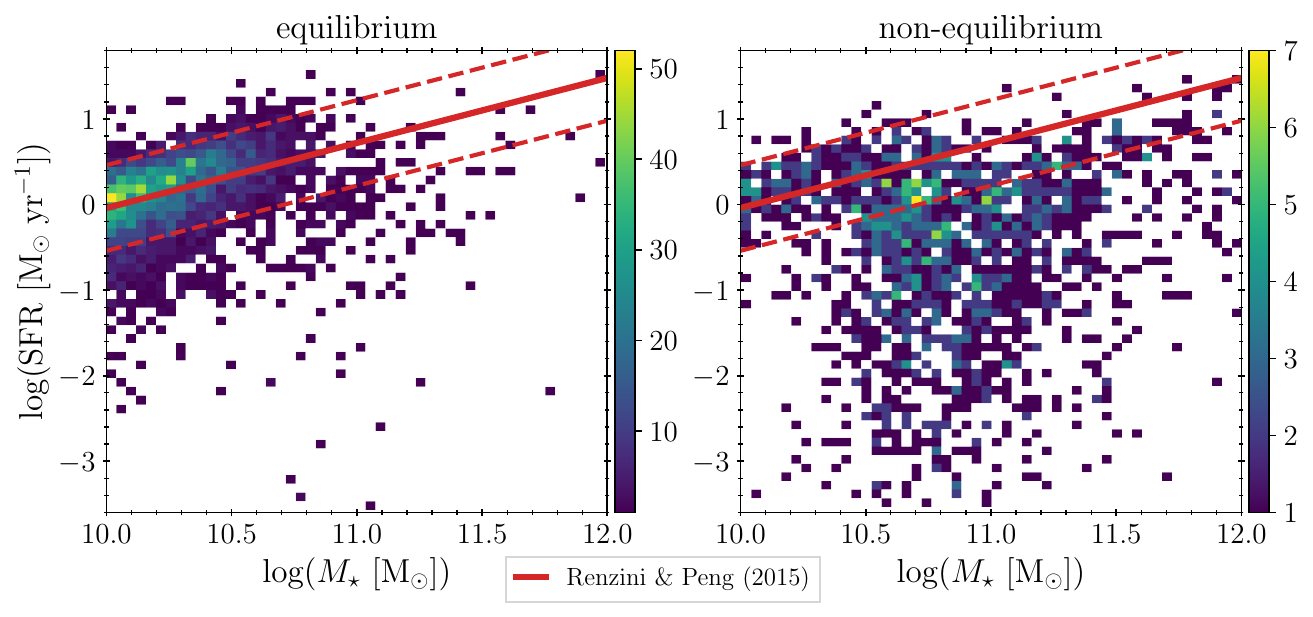}
\caption{SFR as a function of stellar mass for equilibrium (left) and non-equilibrium (right) galaxies. The solid red line indicates the SFMS from \citet{Renzini+Peng2015ApJ...801L..29R}, with dashed lines showing the adopted scatter. Colour maps indicate the number of galaxies per bin in each panel; note that the colour scales differ between the equilibrium and non-equilibrium panels, reflecting the different sample sizes.}
\label{fig:SFRmainseq}
\end{figure*}

To examine whether chemical decoupling is associated with distinct star-formation behaviour, Fig.~\ref{fig:SFRmainseq} shows the distribution of equilibrium (left) and non-equilibrium (right) galaxies in the $M_\star-\mathrm{SFR}$ plane. The solid red line indicates the ridge of the star-formation main sequence (SFMS) for the local Universe \citep{Renzini+Peng2015ApJ...801L..29R}. A clear dichotomy is visible. Equilibrium galaxies predominantly lie within the scatter of the SFMS, exhibiting SFRs consistent with typical star-forming systems at fixed stellar mass.

In contrast, non-equilibrium galaxies display a systematic offset towards lower SFRs, placing many of them in a partially quenched or transitional regime. This indicates that chemical decoupling is closely linked to suppressed star formation in massive galaxies. The reduced star-formation activity implies that, while these systems retain significant stellar mass and metal-rich stellar populations, their present-day gaseous component is less efficiently forming new stars.

This distinct combination of high stellar metallicity, diluted gas-phase metallicity, and suppressed star formation is consistent with a scenario in which non-equilibrium galaxies represent a late evolutionary phase following merger activity or accretion episodes. In such systems, gas inflows can dilute the interstellar medium, while feedback and morphological transformation act to stabilise the gas against further star formation \citep[e.g.][]{Martig+2009ApJ...707..250M}, delaying the recovery of chemical equilibrium. Crucially, the resulting suppression of star formation halts the enrichment cycle, locking the galaxy in a chemically decoupled state for an extended period.

\subsection{Gas fractions and future star-formation potential}\label{sec:gas_fraction}

\begin{figure}[!h]
\centering \includegraphics[width=\columnwidth, keepaspectratio]{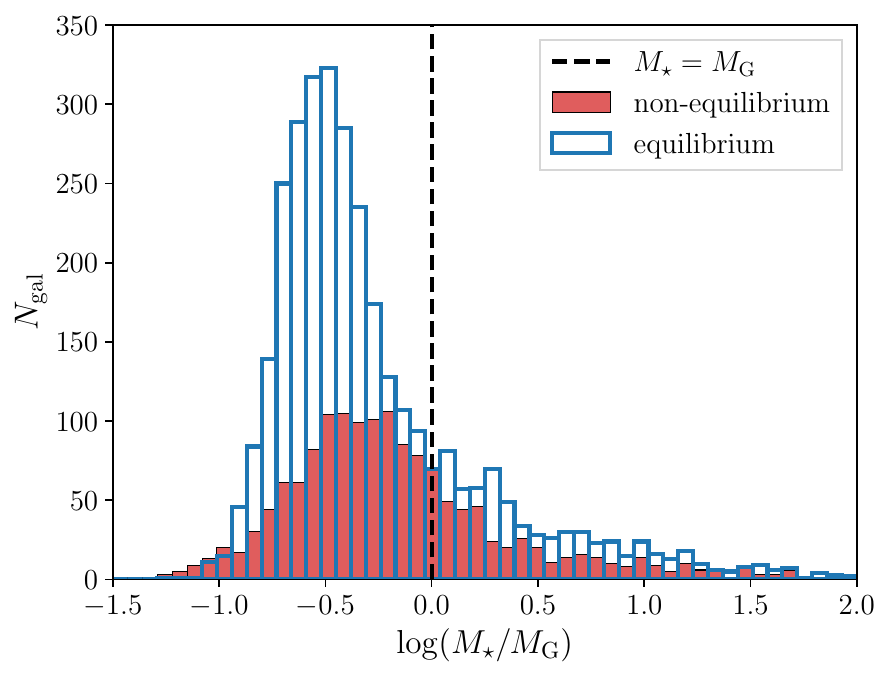}
\caption{Distribution of the stellar-to-gas mass ratio, $\log(M_\star/M_\mathrm{G})$, for equilibrium (blue) and non-equilibrium (red) galaxies in the massive, star-forming sample. The vertical dashed line marks equal stellar and gas mass, as indicated in the legend. Histograms show the number of galaxies in each population as a function of the stellar-to-gas mass ratio.} 
\label{fig:gasfraction}
\end{figure}

Further insight into the evolutionary state of chemically decoupled galaxies is provided by their gas content. Figure~\ref{fig:gasfraction} shows the distribution of the stellar-to-gas mass ratio for equilibrium and non-equilibrium systems. The vertical dashed line ($M_\star = M_\mathrm{G}$) marks the transition from gas-rich to stellar-dominated regimes. The equilibrium population (blue) is predominantly gas-rich, consistent with their ongoing activity on the SFMS. In contrast, the non-equilibrium galaxies (red) exhibit a pronounced tail that extends into the gas-poor regime, i.e. $\log (M_\star/M_\mathrm{G})>0$, corresponding to lower gas fractions relative to their stellar mass.

It is important to note that non-equilibrium galaxies are also systematically more massive than equilibrium systems, so their total gas masses may remain comparable. The observed difference, therefore, reflects primarily a reduced gas fraction rather than an absence of gas. Nevertheless, lower gas fractions indicate a diminished capacity for sustained star formation and chemical enrichment relative to the existing stellar mass, consistent with the partially quenched nature of these galaxies.

Taken together, our results demonstrate that chemical decoupling in massive galaxies is associated with both suppressed star formation and reduced gas fractions, supporting the view that these systems occupy a transitional evolutionary stage in which the enrichment histories of stars and gas have reached a temporal offset. The systematically reduced gas fractions of non-equilibrium galaxies, combined with their suppressed SFRs, place these systems in a regime broadly analogous to the observationally defined green valley population, commonly interpreted as a transitional phase between actively star-forming and quiescent galaxies. This is consistent with observational studies showing that green valley galaxies exhibit reduced molecular and atomic gas fractions relative to main-sequence systems of comparable stellar mass \citep[e.g.][]{Lin+2022ApJ...926..175L, Lin+2024ApJ...963..115L, Lin+2026ApJ...999..263L, Colombo+2025A&A...699A.366C}.

\section{Habitability implications}\label{sec:astrobio-results}

The chemical and star-formation differences established in Sect.~\ref{sec:phys-results} naturally raise the question of how chemically decoupled galaxies would be ranked by commonly used galaxy-scale habitability metrics. To explore this, we adopted the framework introduced by \citet{Dayal+2015ApJ...810L...2D}, which estimates the abundance of terrestrial planets in a galaxy by combining its stellar mass ($M_\star$), gas-phase metallicity ($Z$), and SFR. More precisely, this framework estimates the number of habitable planets $N_\mathrm{p}$ as a function of the total stellar mass (representing the number of potential hosts), the gas-phase metallicity (governing present and future planet formation efficiency), and the SFR (serving as a proxy for the sterilisation risk from supernovae and gamma-ray bursts). The relation is expressed as 

\begin{equation}\label{eq:dayal}
    N_\mathrm{p} \propto \frac{M_\star^2 (Z/Z_\odot)^\alpha}{\mathrm{SFR}}
.\end{equation}

\noindent The exponent $\alpha$ parametrises the strength of the metallicity dependence of terrestrial planet formation. Although giant planet formation exhibits a strong dependence on host metallicity ($\alpha \simeq 2$), recent analyses of Kepler data suggest that the occurrence rate of terrestrial planets is consistent with a linear or weaker scaling \citep[e.g.][]{Buchhave+2012Natur.486..375B, Wang+Fischer2015AJ....149...14W}. The metallicity dependence of small planet occurrence remains uncertain and may be weaker than that of giant planets. Observational selection effects compound this uncertainty: exoplanet samples are strongly shaped by detection biases (e.g. transit depth, survey completeness, and host-star properties), making it challenging to infer intrinsic metallicity dependences for small planets, but also the distribution of habitable planets across the Galaxy \citep{2025A&A...704A.275P}. In particular, terrestrial-planet occurrence rates derived from Kepler are subject to completeness corrections and potential biases in the underlying stellar sample \citep[e.g.][]{Howard+2012ApJS..201...15H, Gaidos+Mann2014ApJ...791...54G, Winn+Fabrycky2015ARA&A..53..409W}. However, a linear prescription provides a useful reference case for comparing relative trends within the simulated galaxy population, and therefore we adopted $\alpha = 1$ in this work. Moreover, our choice of $\alpha = 1$ is conservative; weaker dependence ($\alpha < 1$) would result in even higher values of the relative habitability proxy. It is important to emphasise that the proxy defined in Eq.~\ref{eq:dayal} traces the present-day abundance of potentially habitable planets, and as it is dominated by the existing stellar population and the current rate of sterilising events, it does not explicitly model the time evolution of planet formation. However, the same physical quantities entering the proxy, in particular the gas-phase metallicity ($Z$) and SFR, also regulate the capacity of galaxies to form new planetary systems. Lower gas-phase metallicities reduce the efficiency of terrestrial planet formation, whereas the SFR provides a limit for the production of new stellar hosts. Consequently, although not directly encoded in Eq.~\ref{eq:dayal}, these quantities provide a physically motivated indication of the future potential for planet formation and, as we discuss below, it is precisely the behaviour of these quantities in chemically decoupled systems that gives rise to the habitability trade-off.

A methodological caveat in this analysis is the choice of the metallicity tracer. Standard applications of the \citet{Dayal+2015ApJ...810L...2D} formalism assume chemical equilibrium, using gas-phase metallicity ($Z_\mathrm{G}$) as a proxy for the stellar population. In galaxies where stellar and gas-phase metallicities are systematically offset, using stellar metallicity ($Z_\star$) alone may overestimate the metal content of the present-day interstellar medium and, therefore, misrepresent the conditions relevant for ongoing planet formation. For this reason, and to maintain consistency with the \citet{Dayal+2015ApJ...810L...2D} framework, we evaluated the habitability proxy using the gas-phase metallicity ($Z_\mathrm{G}$), which more directly reflects the chemical state of the current star-forming reservoir. However, as demonstrated in Sect.~\ref{sec:equil_and_non_equil}, the non-equilibrium population is defined by a systematic deficit in gas-phase metallicity relative to the stellar component, that is, $Z_\mathrm{G} < Z_\star$. Consequently, the use of $Z_\mathrm{G}$ in the calculation of $N_\mathrm{p}$ yields a conservative lower limit for these systems. Since the established stellar populations are significantly more metal-rich than the diluted interstellar medium, the true fraction of stars hosting planets is likely higher than our derived values indicate.

\begin{figure}[!h]
\centering \includegraphics[width=\columnwidth, keepaspectratio]{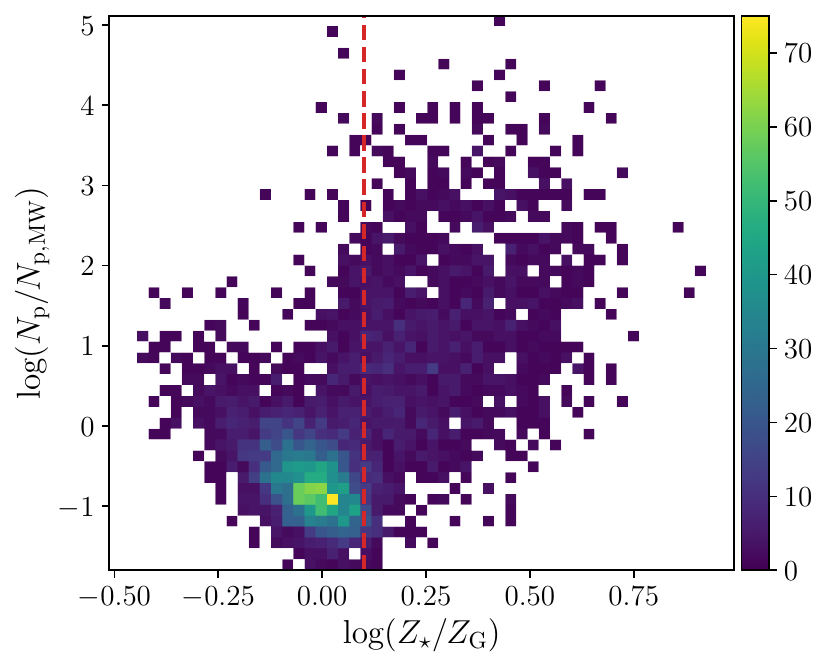}
\caption{Terrestrial planet number proxy \citep{Dayal+2015ApJ...810L...2D} relative to the MW, $N_\mathrm{p}/N_\mathrm{p, MW}$, as a function of the stellar-to-gas metallicity ratio for galaxies in the final sample. Colours indicate the number of galaxies per bin, and the vertical dashed red line separates equilibrium from non-equilibrium galaxies.}
\label{fig:Dayal}
\end{figure}

To facilitate comparison across the sample, we computed planet numbers relative to the MW, with the resulting terrestrial planet proxy:

\begin{equation}
  \frac{N_\mathrm{p}}{N_{\mathrm{p}, \mathrm{MW}}} = \left(\frac{M_\star}{M_{\star, \mathrm{MW}}}\right)^2 \left( \frac{Z_\mathrm{G}}{Z_\mathrm{MW}}\right)^\alpha \left(\frac{\mathrm{SFR_{MW}}}{\mathrm{SFR}}\right)  
\end{equation}

\noindent adopting standard constants and values for the MW commonly used in the literature \citep[e.g.][]{Dayal+2015ApJ...810L...2D, Stojkovic+2019MNRAS}. Specifically, we used the stellar mass $M_{\star,\mathrm{MW}} = 6 \times 10^{10}\; \mathrm{M_\odot}$ and $\mathrm{SFR_{MW} \simeq 3\; \mathrm{M_\odot} \mathrm{yr}^{-1}}$ \citep{Licquia+Newman2015ApJ...806...96L}. The solar metallicity value of $Z_\mathrm{MW} = 0.0196$, used here as a proxy for the present-day gas-phase metallicity of the MW, is taken from \citet{Vagnozzi+2017ApJ...839...55V}, following the recipe from \citet{Stojkovic+2019MNRAS}. We emphasise that this proxy should be interpreted as a relative indicator rather than an absolute prediction of planet counts, and is most useful for identifying systematic differences between different galaxy populations identified in this work. Figure~\ref{fig:Dayal} shows the resulting distribution of this proxy as a function of the stellar-to-gas metallicity ratio $Z_\star/Z_\mathrm{G}$, with the dashed red vertical line separating equilibrium and non-equilibrium galaxy populations.

Despite this conservative estimate, which was previously noted, Fig.~\ref{fig:Dayal} reveals a substantial positive correlation between chemical decoupling and relative habitability. A clear systematic trend is apparent: galaxies with a larger chemical imbalance, characterised by elevated $Z_\star/Z_\mathrm{G}$, typically exhibit higher values of $N_\mathrm{p}/N_\mathrm{p, MW}$. The visual trend observed in Fig.~\ref{fig:Dayal} is substantiated by a quantitative comparison of the habitability distributions for the two populations. The equilibrium galaxies exhibit a median habitability score of $N_\mathrm{p}/N_\mathrm{p, MW} \simeq 0.22$, indicating that the typical system in chemical equilibrium possesses a current habitability potential lower than that of MW. In stark contrast, the non-equilibrium population displays a median score of $N_\mathrm{p}/N_\mathrm{p, MW} \simeq 8.57$. This corresponds to an approximately 40-fold increase in the habitability of the typical chemically decoupled galaxy relative to its equilibrium counterpart. Although both populations include rare extreme outliers, chemically decoupled systems systematically dominate the high-$N_\mathrm{p}$ tail, with maximum values reaching $\sim 1.3\times 10^5$ relative to the MW. These statistics quantitatively reinforce the trend shown in Fig.~\ref{fig:Dayal}, demonstrating that chemical decoupling is associated with a strong enhancement in the inferred proxy of present-day habitability.

This behaviour is driven by the combined effect of the three parameters entering the scaling of \citet{Dayal+2015ApJ...810L...2D}. Non-equilibrium galaxies are systematically more massive than equilibrium systems, boosting the proxy through the strong $M_\star^2$ dependence. At the same time, these galaxies exhibit suppressed SFRs relative to SFMS (Fig.~\ref{fig:SFRmainseq}, partial quenching), which further increases $N_\mathrm{p}$ through the inverse SFR dependence. As a consequence, chemically decoupled galaxies dominate the high-$N_\mathrm{p}$ tail of the distribution despite representing only $\sim 30\%$ of the massive star-forming population. In the context of this habitability prescription, chemically decoupled systems therefore correspond to environments of enhanced present-day habitability, reflecting both their large established stellar populations and their reduced incidence of ongoing energetic feedback events.

\begin{figure}[!h]
\centering
\includegraphics[width=\columnwidth, keepaspectratio]{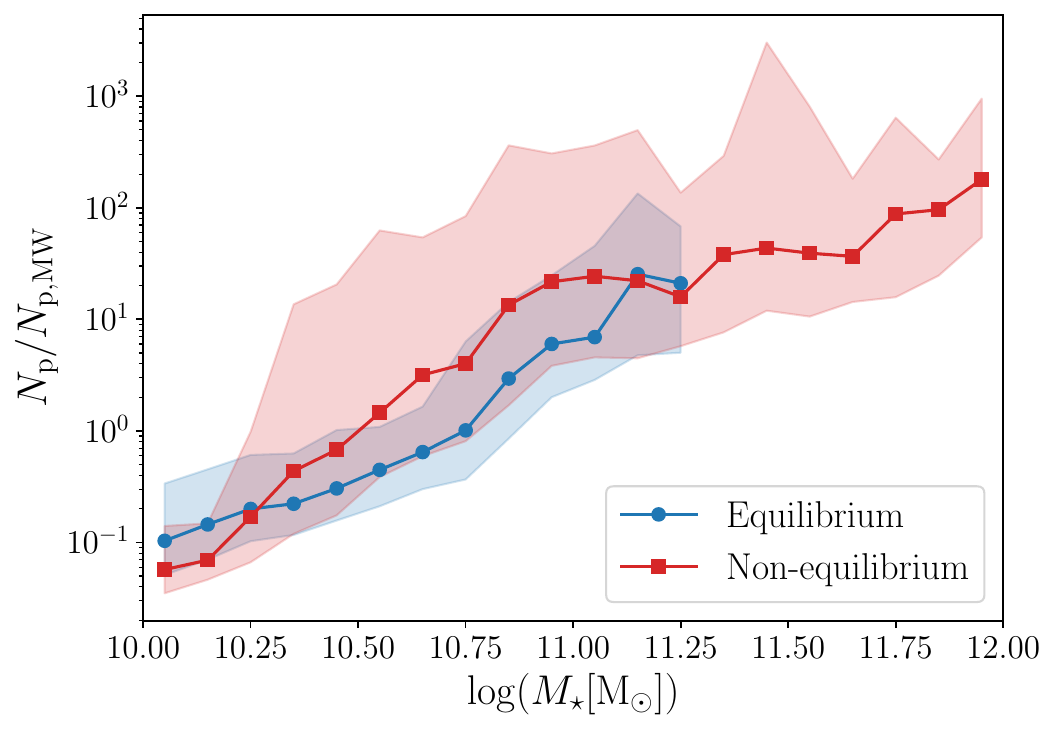}
\caption{Median terrestrial planet proxy $N_{\mathrm{p}}/N_{\mathrm{p, MW}}$, normalised to the MW, as a function of stellar mass for equilibrium (blue circles) and non-equilibrium (red squares) galaxies. Shaded regions indicate the 16th–84th percentile range in each mass bin. Only bins containing at least ten galaxies per population are shown.}
\label{fig:Nt_massbins}
\end{figure}

Since non-equilibrium galaxies are systematically more massive, to assess whether the enhancement in $N_\mathrm{p}/N_\mathrm{p, MW}$ arises solely from the higher stellar masses of the non-equilibrium population, we examined the behaviour of the terrestrial planet proxy at fixed stellar mass. Figure~\ref{fig:Nt_massbins} shows the median $N_\mathrm{p}/N_\mathrm{p, MW}$ as a function of stellar mass for equilibrium and non-equilibrium galaxies separately. Although the proxy increases with stellar mass for both populations, non-equilibrium galaxies exhibit systematically higher median values at fixed mass above $\log_{10}(M_\star[\mathrm{M_\odot}]) \sim 10.5$. The median offset reaches factors of approximately $2$–$4$ in the intermediate- and high-mass regime. This demonstrates that the enhancement in the habitability proxy cannot be attributed purely to stellar mass selection but instead reflects the combined influence of suppressed star formation and chemical decoupling in these systems. However, we emphasise that chemical decoupling acts as an indirect driver of the habitability proxy, primarily through its impact on star formation and gas-phase metallicity, rather than as an independent parameter.

\begin{figure}
    \centering
    \includegraphics[width=\linewidth, keepaspectratio]{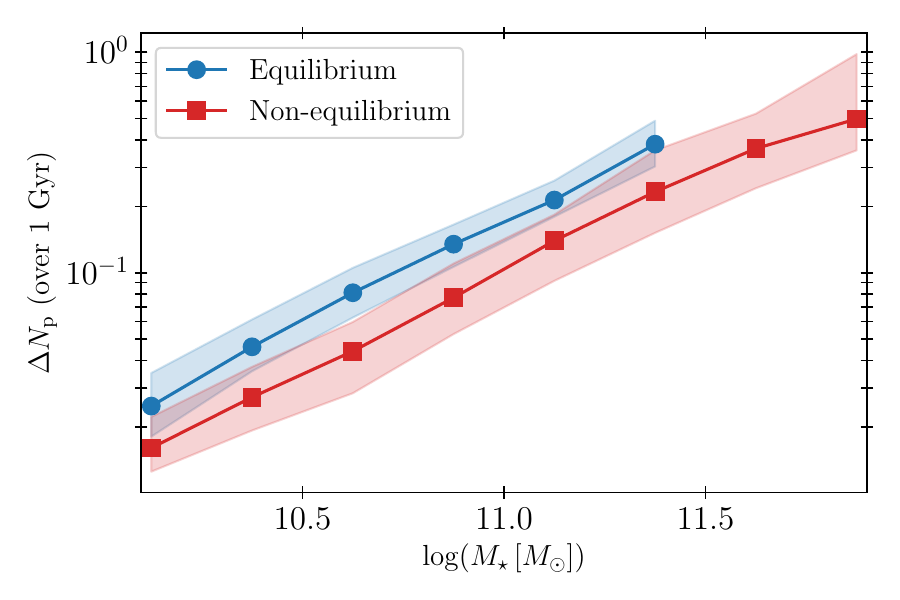}
    \caption{Illustrative estimate of the increment in the terrestrial planet proxy, $\Delta N_{\rm p}$, accumulated over $1\;\textrm{Gyr}$ assuming that both the SFR and gas‑phase metallicity ($Z_\mathrm{G}$) remain constant at their present‑day values, as a function of stellar mass ($M_\star$) for equilibrium (blue circles) and non-equilibrium (red squares) galaxies. Shaded regions indicate the 16th–84th percentile range in each mass bin. Only bins containing at least ten galaxies per population are shown.}
    \label{fig:delta_np}
\end{figure}

To further assess the implications of chemical decoupling for future terrestrial planet formation, we performed a simple illustrative calculation within the \citet{Dayal+2015ApJ...810L...2D} framework. For each galaxy in the sample, we computed the expected increment in the planet proxy, $\Delta N_\mathrm{p}$, accumulated over a period of $1\; \textrm{Gyr}$, under the conservative assumption that the SFR and the gas-phase metallicity ($Z_\mathrm{G}$) remain constant at their present-day value. This assumption is deliberately optimistic for non-equilibrium galaxies, since partially quenched systems are expected to experience further SFR decline over time; any realistic SFR evolution would therefore amplify the effect described here. The resulting $\Delta N_\mathrm{p}$ captures the contribution of newly formed stellar populations to the total planet count, and is governed not only by the rate of star formation but also by the gas-phase metallicity $Z_\mathrm{G}$, which sets the terrestrial planet formation efficiency per unit of new stellar mass within the framework of Eq.~\ref{eq:dayal}. Figure~\ref{fig:delta_np} shows the median $\Delta N_\mathrm{p}$, with the 16th–84th percentile range indicated by the shaded regions, as a function of stellar mass for both populations. Although $\Delta N_\mathrm{p}$ increases with stellar mass for both equilibrium and non-equilibrium galaxies, non-equilibrium galaxies exhibit systematically lower $\Delta N_\mathrm{p}$ at fixed stellar mass across the full mass range considered. We note that the choice of $1\; \textrm{Gyr}$ is arbitrary, as $\Delta N_\mathrm{p}$ scales linearly with the assumed time interval; any other choice would yield a qualitatively identical result. This calculation illustrates how the reduced future planet formation efficiency of non-equilibrium galaxies follows directly from their present-day chemical properties.

Although the framework utilised in this work highlights chemically decoupled (i.e. non-equilibrium) galaxies as highly favourable in terms of present-day habitability, their physical properties simultaneously suggest a diminished or limited capacity for future terrestrial planet formation compared to the equilibrium galaxies of the same stellar mass, as previously shown. This interpretation follows from the same observable quantities already characterising these systems in Sect.~\ref{sec:phys-results}, interpreted within the physical context of the \citet{Dayal+2015ApJ...810L...2D} framework. First, the rate at which new terrestrial planets can form in a galaxy scales with its SFR, since new planets emerge around newly formed stars. Non-equilibrium galaxies are systematically offset below the SFMS (Fig.~\ref{fig:SFRmainseq}), meaning they are forming fewer new stars per unit time than equilibrium galaxies of the same stellar mass. A lower SFR therefore implies a reduced rate of new planetary system formation, irrespective of other galaxy properties. Second, the efficiency of terrestrial planet formation around any newly formed stars is expected to scale with the metallicity of the natal gas reservoir (here traced by the gas-phase metallicity $Z_\mathrm{G}$), as previously mentioned. Non-equilibrium galaxies exhibit systematically lower $Z_\mathrm{G}$ than equilibrium counterparts of comparable stellar mass, which is the defining characteristic of chemical decoupling explored in this work. Stellar populations forming from this diluted interstellar medium are therefore expected to do so with reduced planet-formation efficiency relative to equilibrium systems.

Taken together, these effects suggest that not only are fewer new stars forming in non-equilibrium galaxies, but those that do form originate from a chemically diluted reservoir, further reducing their individual planet-hosting potential. In this sense, the rate of terrestrial planet production is suppressed relative to what would be expected for an equilibrium system of the same stellar mass. This leads to a habitability trade-off that is inherently time-dependent: the physical conditions that maximise the present-day inventory of potentially habitable worlds (low SFRs, reducing sterilising events, and high established stellar masses) are also those that minimise the rate at which new habitable worlds are formed. Chemically decoupled galaxies can therefore be understood as systems experiencing a transient peak in habitable potential rather than a sustained state of enhanced habitability.

\begin{figure*}[!h]
\centering \includegraphics[width=2\columnwidth, keepaspectratio]{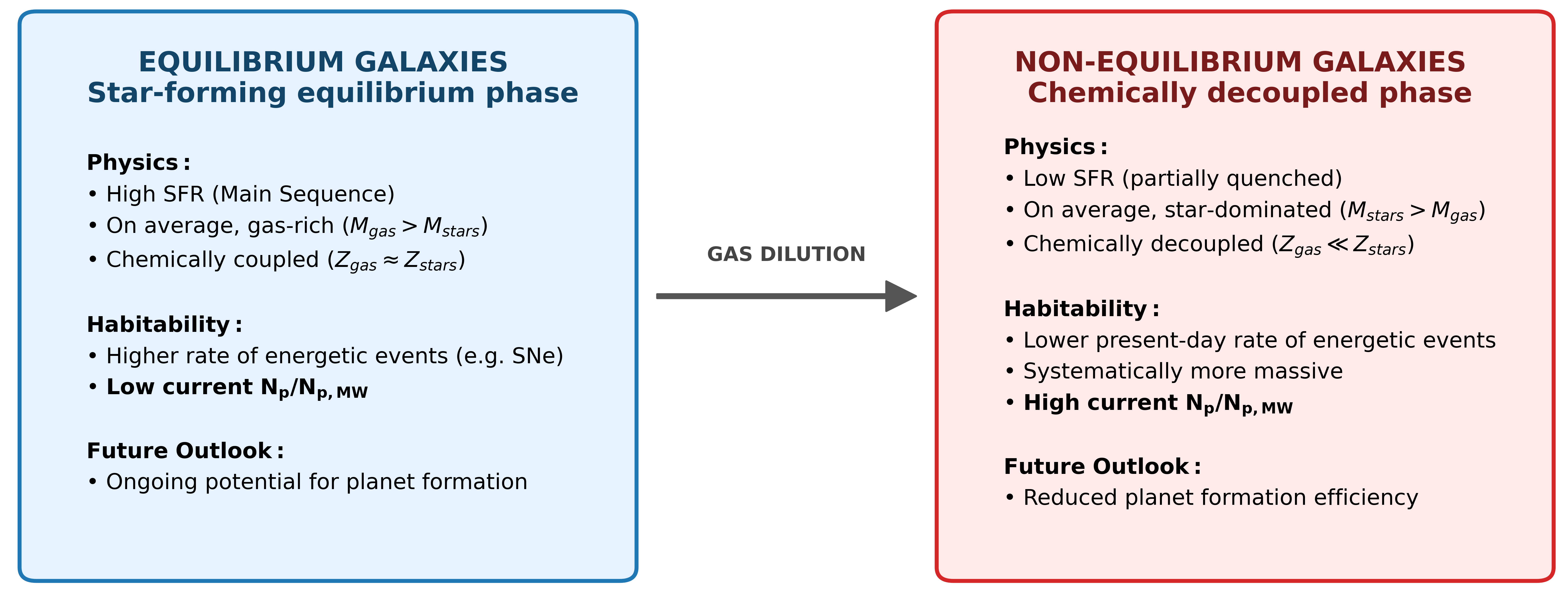}
\caption{Schematic summary of the two galaxy populations identified in this work. This diagram is intended as a qualitative interpretation of the evolutionary phase associated with chemical decoupling.}
\label{fig:schematic}
\end{figure*}

In this sense, chemically decoupled galaxies may represent a late evolutionary stage in which the existing population of potentially habitable planets resides in a relatively low-hazard environment, while the renewal of habitable worlds through new planet formation proceeds less efficiently than in systems in chemical equilibrium of comparable stellar mass. More broadly, this interpretation suggests that galactic habitability is not a static property determined solely by stellar mass and metallicity, but instead depends sensitively on the chemical state of the gas reservoir and the evolutionary phase of the host galaxy. Chemically decoupled massive galaxies can therefore be understood as systems that are highly favourable for habitability at the present epoch, yet less capable of sustaining this state over cosmic timescales. A schematic summary of this evolutionary interpretation is provided in Fig.~\ref{fig:schematic}.

\section{Discussion}\label{sec:discussion}

The results presented in Sects. \ref{sec:phys-results} and \ref{sec:astrobio-results} suggest that chemically decoupled galaxies represent a distinct evolutionary stage in the massive galaxy population, characterised by metal-rich stellar components, diluted gas-phase metallicities, suppressed star formation, and typically reduced gas fractions. This configuration is naturally produced by processes that affect the interstellar medium more rapidly than the established stellar population \citep[e.g.][]{Yates+2012MNRAS.422..215Y, Peng+Maiolino2014MNRAS.438..262P}, including merger-driven inflows \citep[e.g.][]{Bustamante+2018MNRAS.479.3381B, Sparre+2022MNRAS.509.2720S}, external gas accretion \citep[e.g.][]{Hwang+2019ApJ...872..144H, Pan+2025ApJ...982..130P}, and quenching-related stabilisation of the gas reservoir \citep{Martig+2009ApJ...707..250M}. Both semi-analytic models and hydrodynamical simulations have shown that mergers and accretion episodes can drive significant metallicity dilution in the gas phase while leaving the stellar metallicity largely unchanged \citep{Yates+2012MNRAS.422..215Y, Yates+2014MNRAS.439.3817Y, Bustamante+2018MNRAS.479.3381B}. In this context, chemical decoupling is expected to be most prominent in massive systems transitioning away from sustained star formation.

Although this study is based on the IllustrisTNG simulation, M31 offers a local observational analogue to the non-equilibrium population identified in our sample. Physically, M31 is different from MW in terms of its assembly history and current dynamical state. It is a more massive system \citep[e.g. $\log(M_\star/\mathrm{M_\odot}\approx 11)$,][]{Sick+2015IAUS..311...82S} that resides significantly below the SFMS, with a global SFR estimated at roughly $0.2-0.4 \; \mathrm{M_\odot} \mathrm{yr}^{-1}$ \citep{Barmby+2006ApJ...650L..45B, Tabatabaei+Berkhuijsen2010A&A...517A..77T, Ford+2013ApJ...769...55F, Rahmani+2016MNRAS.456.4128R}. This combination of high stellar mass and reduced SFR mirrors the partially quenched nature of chemically decoupled galaxies in our simulation sample.

M31 also shows clear evidence of a complex accretion and interaction history, which may have contributed to perturbations and dilution of its gaseous reservoir. The Pan-Andromeda Archaeological Survey (PAndAS) has provided a panoramic view of the Andromeda halo, revealing a wealth of substructures (\citealp{Ibata2007, Ibata2014, Martin2016}). The Giant Stellar Stream was discovered by \citet{Ibata2001} and the shell system by \citet{Ferguson2002}. These structures are a result of the accretion of a satellite galaxy. Many observations were made to determine the metallicity in these structures. Chemical abundances and kinematics are given in SPLASH (\citealp{Guhathakurta, Kalirai2006, Gilbert2009, Gilbert2014}) and spectroscopic analyses in the Elemental Abundances in M31 collaborations (\citealp{Gilbert2019, Escala2020a, Escala2020b}). Deep surveys have revealed an extensive tidal substructure in M31's halo, consistent with significant merger and satellite accretion events over its recent evolutionary history \citep{McConnachie+2009Natur.461...66M}. Such interactions provide a natural channel for introducing lower-metallicity gas into the disc and for driving inflows that dilute the interstellar medium.  Furthermore, detailed theoretical studies of the stellar substructures in the halo and disc of M31 indicate a violent accretion history, including a major merger event approximately a couple of gigayears ago \citep[e.g.][]{Hammer+2018MNRAS.475.2754H, DSouza+Bell2018NatAs...2..737D}. Several recent studies continue to refine M31's merger history and chemo-dynamical properties of its substructures \citep[e.g.][]{Dey+2023ApJ...944....1D, Milosevic+2022MNRAS.511.2868M, Milosevic+2024MNRAS.527.4797M, Lewis+2023MNRAS.518.5778L, Tsakonas+2025A&A...699A..56T}.

Recent spectroscopic analyses of planetary nebulae have identified chemically distinct accretion features in the M31 disc, consistent with the presence of metal-poor gas accretion and spatially inhomogeneous enrichment \citep{Arnaboldi+2022A&A...666A.109A, Bhattacharya+2022MNRAS.517.2343B, Bhattacharya+2023MNRAS.522.6010B, Kobayashi+2023ApJ...956L..14K}. The coexistence of younger, relatively metal-poor stellar components with an older, more metal-rich disc population suggests that the enrichment histories of stars and gas in M31 have become partially decoupled. Moreover, \citet{Wojno+2023ApJ...951...12W} provided additional spectroscopic evidence that the formation and merger history of M31 is vastly different compared to our own Galaxy, suggesting that the evolutionary pathways relevant for habitability also differ accordingly. Although direct global measurements of stellar-to-gas metallicity ratios remain challenging observationally, M31 therefore provides a compelling nearby analogue supporting the plausibility of the dilution-driven chemical decoupling that temporarily boosts inferred present-day habitability, as explored in this work.

Although the GHZ of the MW has been extensively explored, M31 has also been considered in a small number of studies on chemical evolution and habitability. In particular, \citet{Carigi+2013}, \citet{Spitoni+2014}, and \citet{2025RLSFN..36..739S} extended the GHZ modelling to Andromeda and suggested that, due to its star-formation history and metallicity gradient, M31 may host an unusually broad spatial region capable of supporting terrestrial planet formation. In these models, the combination of a massive, metal-rich disc and an extended high-metallicity reservoir could, at least qualitatively, imply a larger absolute number of potentially habitable systems than in the MW. However, such inferences remain highly model-dependent and should be regarded as speculative. Comparisons between MW and M31 are sensitive to assumptions regarding chemical enrichment pathways, the temporal evolution of star formation, and the redistribution of stars via radial migration \citep[e.g.][]{2023PASA...40...54M, Spitoni+2025A&A...700A..58S}. Furthermore, net habitability critically depends on the relative importance of metallicity versus sterilising feedback mechanisms \citep{Gowanlock+2011}. At the present epoch, M31's suppressed star-formation activity could, in principle, imply a reduced rate of ongoing energetic events compared to the MW, potentially favouring relatively benign conditions for the long-term survival of existing terrestrial biospheres. In this sense, galaxies such as M31 may resemble the chemically decoupled systems identified in this work, which exhibit enhanced inferred present-day habitability under prescriptions such as \citet{Dayal+2015ApJ...810L...2D}. However, any such present-day advantage must be viewed with caution and as a transient feature. The substantial merger and accretion history of M31 suggests that its interstellar environment may have been considerably more hostile in the past, with elevated sterilisation hazards capable of disrupting planetary atmospheres and biospheres over cosmic time \citep[e.g.][]{Lineweaver+2004}. Thus, while M31 may plausibly host a large absolute population of terrestrial planets, the degree to which these environments have remained continuously habitable throughout its evolutionary history remains uncertain. More generally, our results suggest that galactic habitability is a transient and time-dependent evolutionary property shaped by both chemical and dynamical histories, rather than a static outcome determined solely by present-day stellar mass and metallicity.

Moreover, the long-term stability of habitable environments in both galaxies depends on their future dynamical evolution within the Local Group. While M31 has historically been discussed as a future merger partner of the MW, recent astrometric constraints and dynamical modelling have highlighted that the timing and even the inevitability of such a merger remain uncertain \citep{Salomon+2021MNRAS.507.2592S, Font2025NatAs...9.1107F, Sawala+2025NatAs...9.1206S}. This further emphasises that habitability should be viewed as an evolving, time-dependent property rather than a static outcome. We emphasise that this time-dependent interpretation should be understood as a physically motivated extension of the proxy employed in this work, rather than a direct prediction of its temporal evolution.

\section{Summary and conclusions}\label{sec:conslusions}

In this work we investigated the chemical and star-formation properties of massive galaxies in the IllustrisTNG (TNG100) simulation, and explored the implications of chemical decoupling between stars and gas for galaxy-scale habitability metrics using the standard prescriptions from \citet{Dayal+2015ApJ...810L...2D}.

Our main results can be summarised as follows:
\begin{itemize}
    \item Massive galaxies exhibit systematic stellar–gas chemical decoupling. Independent fits of the stellar and gas-phase MZRs at $z=0$ reveal that the gas-phase metallicity saturates at a lower level than the stellar metallicity, producing a population of galaxies in which $Z_\mathrm{G} < Z_\star$.
    \item Chemically decoupled (i.e. non-equilibrium) galaxies are common among massive star-forming or partially quenched systems. Using the stellar-to-gas metallicity ratio as an operational diagnostic, we identified 1487 non-equilibrium galaxies out of a final sample of 4718 massive star-forming systems, corresponding to $\sim 31.5\%$ of the examined population.
    \item Chemical decoupling is closely related to suppressed star formation and reduced gas fractions. Chemically decoupled galaxies are systematically offset below the SFMS and typically exhibit lower gas fractions, consistent with a transitional evolutionary stage associated with dilution, accretion, or partial quenching, typically in post-merger systems.
    \item Chemical decoupling strongly modifies the inferred present-day habitability. Applying a terrestrial planet abundance proxy that combines the stellar mass, gas-phase metallicity, and SFR, we find that non-equilibrium galaxies dominate the high end of the inferred present-day habitability distribution. This is driven primarily by their high stellar masses and reduced ongoing star formation. Chemical decoupling should therefore be interpreted as a tracer of this transitional evolutionary phase rather than a direct driver of enhanced habitability.
    \item Galactic habitability is intrinsically time-dependent. Although chemically decoupled systems appear favourable at the present epoch, their diluted gas reservoirs and diminished star formation imply a reduced capacity for continued terrestrial planet formation through two compounding effects: the suppressed SFR limits the rate of new planetary system formation, whilst reduced gas-phase metallicity lowers the terrestrial planet-formation efficiency of any stars that do form. Such galaxies may therefore evolve towards increasingly fossilised populations of existing habitable worlds. We interpret this as a habitability trade-off: for galaxies of comparable stellar mass, the same properties that boost the habitability proxy at the present day do so at the expense of future terrestrial planet formation efficiency.
\end{itemize}

Nearby massive galaxies such as M31 exhibit several qualitative similarities to the chemically decoupled (non-equilibrium) population identified here, indicating that chemical decoupling may represent a common late evolutionary phase among MW-mass galaxies. More broadly, our results highlight that galactic habitability cannot be fully captured by present-day stellar mass and stellar metallicity alone: the chemical state of the gas reservoir and the evolutionary stage of the host galaxy play a critical role in shaping both the current and future habitable potential. Chemical decoupling, therefore, reframes galactic habitability as a transient phase of cosmic evolution rather than a static property of galaxies. Future spatially resolved observations and next-generation simulations will be crucial for testing this picture.

\begin{acknowledgements}
The authors thank the anonymous referee for a careful and constructive report that improved the clarity and depth of the manuscript, and the IllustrisTNG team for making their simulations publicly available. This research was supported by the Ministry of Science, Technological Development and Innovation of the Republic of Serbia (MSTDIRS) through contract no. 451-03-33/2026-03/200002, made with the Astronomical Observatory (Belgrade, Serbia), and contract no. 451-03-33/2026-03/200104, made with the Faculty of Mathematics, University of Belgrade.
\end{acknowledgements}

\bibliographystyle{aa} 
\bibliography{aa59550-26}

\end{document}